\documentclass[12pt]{article}
\usepackage{epsfig,axodraw}
\usepackage{axodraw}

\newcommand{\beq}{\begin{equation}}
\newcommand{\eeq}{\end{equation}}
\newcommand{\beqa}{\begin{eqnarray}}
\newcommand{\eeqa}{\end{eqnarray}}
\newcommand{\beqar}{\begin{eqnarray*}}
\newcommand{\eeqar}{\end{eqnarray*}}
\newcommand\cT{T}

\newcommand{\al}{\alpha}
\newcommand{\be}{\beta}

\def\spa          {\ \ \ }
\def\non          {\nonumber}
\def\ha           {\mbox{$\frac{1}{2}$}}

\def\spa          {\ \ \ }
\def\mand         {\spa\mbox{and}\spa}

\def\Tr           {\mbox{\rm Tr}\,}
\def\STr          {\mbox{\rm STr}\,}

\def\cd           {{\cdot}}
\def\ran          {\rangle}
\def\lan          {\langle}
\def\fsH	{H\!\!\!\!/\,}
\newcommand{\del}{\delta}

\newcommand{\eps}{\epsilon}
\newcommand{\ga}{\gamma}
\newcommand{\Ga}{\Gamma}

\newcommand{\inn}{\!\cdot\!}

\newcommand{\lam}{\lambda}

\newcommand{\z}{\zeta}

\newcommand{\ie}{{\it i.e.,}\ }
\newcommand{\labell}[1]{\label{#1}} 
\newcommand{\reef}[1]{(\ref{#1})}
\newcommand\prt{\partial}

\newcommand\cL{{\cal L}}

\newcommand\bD{\bar{D}}

\begin{document}
\baselineskip 18pt%
\begin{titlepage}
\vspace*{1mm}%
\hfill%
\vspace*{13mm}%
\center{{\bf\Large Highly Symmetric D-brane-Anti-D-brane \\
 Effective Actions 
}}
\begin{center}
{Ehsan Hatefi   \small $^{a,b,}$\footnote{ehsan.hatefi@tuwien.ac.at, e.hatefi@qmul.ac.uk, ehsanhatefi@gmail.com, ehsan.hatefi@cern.ch,}}
\vspace*{0.04cm}
\vskip.1in
{ $^{a}$ Institute for Theoretical Physics, TU Wien
\\
Wiedner Hauptstrasse 8-10/136, A-1040 Vienna, Austria},
\vskip.06in
{ $^{b}$ Mathematical Institute, Faculty of Mathematics, Charles University, P-18675, CR}
\vskip.06in
\vspace*{0.01cm}
\end{center}
\begin{center}{\bf Abstract}\end{center}
\begin{quote}

The entire S-matrix elements of four, five and six point functions of D-brane-anti D-brane system are explored. To deal with symmetries of string amplitudes as well as their all order $\alpha'$ corrections we first address a four point function  of one closed string Ramond-Ramond (RR) and  two real  tachyons on the world volume of brane-anti brane system. We then focus on symmetries of string theory as well as universal tachyon expansion to achieve both string and effective field theory of  an RR and three tachyons where the complete algebraic analysis for the whole S-matrix $<V_{C^{-1}}  V_{T^{-1}} V_{T^{0}}V_{T^{0}} >$ was also revealed. Lastly, we employ all the conformal field theory techniques to  $<V_{C^{-1}}  V_{T^{-1}} V_{T^{0}}V_{T^{0}} V_{T^{0}}>$, working out with symmetries of theory and find out the expansion for the amplitude to be able to precisely discover all order singularity structures of  D-brane-anti-D-brane  effective actions of string theory. Various remarks about the so called generalized Veneziano amplitude and new string couplings are elaborated as well.

 \end{quote}
\end{titlepage}

\section{Introduction}

By dealing with unstable branes, one might have some motivations to gain not only more information about  supersymmetry breaking but also explore new couplings on various time dependent backgrounds as well as working out with the properties of different string theories  \cite{Gutperle:2002ai,Lambert:2003zr,Sen:2004nf}. 
Insisting on D-brane anti-D-brane systems, one may reveal Sakai- Sugimoto model \cite{Sakai:2004cn},  the  low energy  physics of some of QCD models and spontaneous symmetry breaking in the appearance of holographic QCD models \cite{Casero:2007ae} where flavour branes are inserted by different parallel branes and anti branes are taken into account within some backgrounds that are dual to colour confined phenomenon. A brane anti-brane configuration can be thought of a probe if $N_f<< N_c$. As it is evident tachyonic strings play the fundamental role in instability of these systems, hence it is crucial to properly keep track of these tachyon modes in both String and Effective Field Theory (EFT) parts.

\vskip.1in

The effective actions that include tachyon modes was found by A.Sen and others in \cite{Sen:1999md,Bergshoeff:2000dq} which could potentially explain afew remarks and phenomena like  the decays of non-BPS branes \cite{Sen:2002in}. One could follow the established argument in \cite{Sen:2004nf}, calrifying how non-BPS EFT has inserted massless states and in particular tachyons. Within some reasonable field content we have also dealt with 
non-BPS actions in \cite{Garousi:2008ge}. One may also talk about tachyon condensation for brane anti brane system \cite{Sen:1998sm} as well.

\vskip.1in

Whenever the distance between brane and anti-brane takes the value of  smaller than string length scale two real tachyon modes would pop in this configuration, thus it made sense to replace them in an EFT and try to employ their dynamics. To do so, one needs to first learn how to embed them in an EFT so that the consistent results with string amplitudes come along. One can point out to a recent paper \cite{Michel:2014lva} on the dynamics of brane anti brane system which puts forward more evidences towards brane actions in the context
of EFT and loop divergences. Other application for this brane-anti-brane system would be
Brane production  \cite{Bergman:1998xv} as well as describing inflation in string theory in the context of KKLT \cite{Dvali:1998pa}.

\vskip.1in

We would like to make contact with scattering amplitudes to fix the ambiguities  that appeared in each order of string theory and the reason for that is simply because not only are we able to discover new couplings based on direct computations but also one fixes for once all the exact coefficients of string corrections in both IIB, IIA.\footnote{ For higher point functions we introduce \cite{Bjerrum-Bohr:2014qwa} and to properly address string corrections we provide these references \cite{Hatefi:2012zh,Hatefi:2012ve}. All order $\alpha'$ BPS corrections as well as a conjecture \cite{Hatefi:2012rx} on universal $\alpha'$ corrections was illustrated that shockingly applied to both non-BPS and supersymmetric cases.} One even can go further and deal with the thermodynamical facets of brane anti brane system, where it is also recognized that at finite temperature this D-brane anti D-brane is being stabilised and can be entirely associated to black holes for which some implications have been devoted to either AdS/CFT or M-theory approach \cite{Hatefi:2012bp}. D-brane-Anti-D-brane system has also been affecting on the stability of KKLT or in Large Volume Scenario and string compactifications \cite{Polchinski:2015bea}. Given a tied relation between D-branes and RR fields 
 \cite{Polchinski:1995mt}  we just point out on the bound states of the branes
\cite{Witten:1995im} as well. Having set the fact that no duality transformation exists for non-BPS branes, one would have to emphasize that the only way of getting exact all order $\alpha'$ corrections of effective couplings in string theory is just through CFT and scattering methods. To notify further comments we just demonstrate  \cite{Hatefi:2012zh}. Last but not least one might head off reading all standard EFT methods for both Wess-Zumino (WZ) and (non)-BPS DBI effective actions that are verified in detail in \cite{Hatefi:2010ik,Myers:1999ps}.


 \vskip.2in

The  paper is organised so that first we warm up with details on 3 and 4 point functions of D-brane anti-D-brane system, basically we fully address the S-matrix of a closed string Ramond-Ramond (RR) field and 2 tachyons  \cite{Kennedy:1999nn} and explain how one does reconstruct all order exact $\alpha'$ corrections in this context.
\vskip.1in

We then move on to observe more hidden symmetries in non-BPS context  and as such we employ all the Conformal Field Theory (CFT) methods to a five point function of
  $<V_{C^{-1}} V_{T^{-1}} V_{T^{0}} V_{T^{0}}>$ in both type IIA and IIB. Given the selection rules \cite{Hatefi:2013yxa}, EFT and the entire algebraic solutions for integrals on 5-point functions we guess an expansion and test our guess with exact  solutions of the integrals related to that amplitude and then produce all the infinite massless/ tachyon poles and come to an agreement with both string and field theory. Eventually we try to address for the first time a six point function of D-brane anti-D-brane system, deriving $<V_{C^{-1}} V_{T^{-1}} V_{T^{0}} V_{T^{0}} V_{T^{0}}>$ S-matrix and given all symmetries of amplitude we discover its expansion. We also show that in a particular limit that is being called soft limit $4k_2.p\rightarrow 1$ the algebraic solutions for all 6-point functions can be found out. It is worth mentioning that within this limit all massless poles of the S-Matrix can be clearly observed and to be regenerated from an EFT argument as well. Lastly, we construct all order higher derivative corrections to four tachyon couplings in the context of brane anti-brane system.

\vskip.1in

\section{Lower order D-brane-Anti-D-brane Effective Actions }

The so called effective actions of a $D_p\bD_p$-brane of IIA(B) may be found out by inserting tachyonic modes in WZ and DBI effective actions. By taking into account just 2 unstable branes of IIB(A) and projecting them out through $(-1)^{F_L}$ operator one derives such an action. In order to simplify the field content, either 2 tachyons and a guage (scalar) fields take part in the action whereas RR will contribute among Chern-Simons or WZ action and other fields act on DBI as follows \cite{Garousi:2007fk}
\beqa
S_{DBI}&=&-\int
d^{p+1}\sigma \Tr\left(V({\cal T})
\sqrt{-\det(\eta_{ab}
+2\pi\alpha'F_{ab}+2\pi\alpha'D_a{\cal T}D_b{\cal T})} \right)\,\,,\labell{nonab} \eeqa 
Trace in \reef{nonab} has to be symmetric for   
$F_{ab},D_a{\cal T}$, 
${\cal T}$ matrices. The definitions of the entire matrices  are given by
\beqa
F_{ab}=\pmatrix{F^{(1)}_{ab}&0\cr 
0&F^{(2)}_{ab}},\,\,
D_{a}{\cal T}=\pmatrix{0&D_aT\cr 
(D_aT)^*&0},\,\, {\cal T}=\pmatrix{0&T\cr 
T^*&0}\,\labell{M12} \eeqa 
 with $F^{(i)}_{ab}=\prt_{a}A^{(i)}_{b}-\prt_{b}A^{(i)}_{a}$ and $D_{a}T=\prt_{a}T-i(A^{(1)}_a-A^{(2)}_a)T$.
 
 \vskip.1in
 
Having made use of the ordinary trace \reef{nonab} gets substituted to A. Sen 's action \cite{Sen:2003tm} if one makes all kinetic terms symmetrized and evaluates the trace, however, it is shown by CFT and scattering amplitudes in \cite{Hatefi:2012cp,Garousi:2007fk} that Sen's effective action does not produce consistent result  with string amplitudes. Tachyon 's potential in the context of type II scattering amplitude is shown by
 \beqa
V(|T|)&=&1+\pi\alpha'm^2|T|^2+
\frac{1}{2}(\pi\alpha'm^2|T|^2)^2+\cdots
\non\eeqa 
 where $m_{T}^2=-1/(2\alpha')$ and  $T_{p}$ is the
tension of a p-brane. The expansion is also made consistent result  with
potential of $V(|T|)=e^{\pi\alpha'm^2|T|^2}$ that came from BSFT as well \cite{Kutasov:2000aq}.
 
  We argued in \cite{Garousi:2007fk} that just  above effective action can entirely and consistently generate all infinite singularities as well as contact terms of sting theory.
  Indeed we also dicussed that the presence of new mixing couplings like
    $F^{(1)}\cdot{F^{(2)}}$ as well as $D\phi^{(1)}\cdot{D\phi^{(2)}}$ is necessary to derive the actual and consistent results that are matched string computations with EFT. The Lagrangian of SYM (at quadratic level) in the presence of tachyons must be modified and has to have the following structures/ interactions \cite{Garousi:2007fk}: 
\beqa {\cal
L}_{DBI}&\!\!\!=\!\!\!&-T_p(2\pi\alpha')\left(m^2|T|^2+DT\cdot(DT)^{*}-\frac{\pi\alpha'}{2}
\left(F^{(1)}\cdot{F^{(1)}}+
F^{(2)}\cdot{F^{(2)}}\right)\right)+T_p(\pi\alpha')^3\nonumber\\
&&\times\left(\frac{2}{3}DT\cdot(DT)^{*}\left(F^{(1)}\cdot{F^{(1)}}+F^{(1)}\cdot{F^{(2)}}+F^{(2)}\cdot{F^{(2)}}\right)\right.\labell{exp1}\\
&&\left.+\frac{2m^2}{3}|\tau|^2\left(F^{(1)}\cdot{F^{(1)}}+F^{(1)}\cdot{F^{(2)}}+F^{(2)}\cdot{F^{(2)}}\right)\right.\nonumber\\
&&-\left.\frac{4}{3}\left((D^{\mu}T)^*D_{\beta}T+D^{\mu}T(D_{\beta}T)^*\right)\left({F^{(1)}}^{\mu\alpha}F^{(1)}_{\alpha\beta}+{F^{(1)}}^{\mu\alpha}F^{(2)}_{\alpha\beta}+{F^{(2)}}^{\mu\alpha}F^{(2)}_{\alpha\beta}\right)\right)
\nonumber
\eeqa

 WZ action for brane anti brane system  with $C$ becomes a sum on  RR potentials $C=\sum_n(-i)^{\frac{p-m+1}{2}}C_m$ is \cite{Douglas:1995bn} 
\beqa
S = \mu_p\int_{\Sigma_{(p+1)}} C \wedge  \left(e^{i2\pi\alpha'F^{(1)}}-e^{i2\pi\alpha'F^{(2)}}\right)\ ,
\labell{eqn.wz}
\eeqa
In  \cite{Kraus:2000nj} it was justified how to consider the tachyons in the effective actions where another approach would be making contact with superconnection of the non-commutative geometry ~\cite{quil,berl,Roepstorff:1998vh} to be 
\beqa
S_{WZ}&=&\mu_p \int_{\Sigma_{(p+1)}} C \wedge \STr e^{i2\pi\alpha'\cal F}\labell{WZ}\eeqa 
where the curvature and  super-connection are defined accordingly as
\beqa {\cal F}&=&d{\cal A}-i{\cal A}\wedge\cal A\nonumber \eeqa
and 
\begin{displaymath}
i{\cal A} = \left(
\begin{array}{cc}
  iA^{(1)} & \beta T^* \\ \beta T &   iA^{(2)} 
\end{array}
\right) \ ,
\non\end{displaymath}
In \cite{Garousi:2007fk} it is also verified how to derive the curvature as follows
\begin{displaymath}
i{\cal F} = \left(
\begin{array}{cc}
iF^{(1)} -\beta^2 |T|^2 & \beta (DT)^* \\
\beta DT & iF^{(2)} -\beta^2|T|^2 
\end{array}
\right) \ ,
\non\end{displaymath}
with $F^{(i)}=\frac{1}{2}F^{(i)}_{ab}dx^{a}\wedge dx^{b}$ and $DT=[\partial_a T-i(A^{(1)}_{a}-A^{(2)}_{a})T]dx^{a}$. Lastly one does extract various couplings of the WZ action \reef{WZ} to be able to lead to diverse WZ couplings that are needed for consistent result of EFT with string side as follows
\beqa
C\wedge \STr i{\cal F}&\!\!\!\!=\!\!\!&C_{p-1}\wedge(F^{(1)}-F^{(2)})\labell{exp2}\\
C\wedge \STr i{\cal F}\wedge i{\cal F}&\!\!\!\!=\!\!\!\!&C_{p-3}\wedge \left\{F^{(1)}\wedge F^{(1)}-
F^{(2)}\wedge F^{(2)}\right\}\nonumber\\
&& +C_{p-1}\wedge\left\{-2\beta^2|T|^2(F^{(1)}-F^{(2)})+2i\beta^2 DT\wedge(DT)^*\right\}\nonumber\\
C\wedge \STr i{\cal F}\wedge i{\cal F}\wedge i{\cal F}&\!\!\!\!=\!\!\!\!&
C_{p-5}\wedge \left\{F^{(1)}\wedge F^{(1)}\wedge F^{(1)}-
F^{(2)}\wedge F^{(2)}\wedge F^{(2)}\right\}\nonumber\\
&&+C_{p-3}\left\{-3\beta^2|T|^2(F^{(1)}\wedge F^{(1)}-F^{(2)}\wedge F^{(2)})\right.\nonumber\\
&&\left.\qquad \qquad+3i\beta^2(F^{(1)}+F^{(2)})\wedge DT\wedge(DT)^*\right\}\nonumber\\
&&+C_{p-1}\left\{3\beta^4|T|^4\wedge(F^{(1)}-F^{(2)})-6i\beta^4|T|^2DT\wedge (DT)^*\right\}\nonumber
\eeqa

The world volume of a non-BPS brane of both type IIA (IIB) string theory includes a real tachyon and  the three point function of an RR and a tachyon $<V_{C^{-1}} V_{T^{-1}}>$
based on direct CFT methods was given in \cite{Hatefi:2015gwa} to be
\beqa
{\cal A}^{C^{-1}T^{-1}} & \sim & -2i \Tr(P_{-}\fsH_{(n)}M_p)
\label{CT}\eeqa
Using momentum conservation one reveals that  $p^ap_a=k^2=-\frac{1}{4} $, the string amplitude \reef{CT} can be reconstructed out in an EFT by   $
 (2\pi\alpha'\beta'\mu'_p\int C_p\wedge DT)$ coupling where there is no singularity structure for this amplitude at all.
 
 On the other hand, the world volume of a D-brane-anti D-brane system involves two real tachyons. Using direct CFT methods \cite{Friedan:1985ge}  the four point function of an RR and two tachyons $<V_{C^{-1}} V_{T^{-1}} V_{T^{0}}>$ can be explored. To get familiar with notations we just warm up with  this calculation where the vertices with  their CP factors for brane- anti brane are
 \beqa
 V_{T}^{(-1)}(x_1) &=&  e^{-\phi(x_1)}e^{\alpha'ik_1.X(x_1)}\lam\otimes\sigma_2 \nonumber\\
V_{T}^{(0)}(x_2) &=&\alpha'ik_2\cd\psi(x_2) e^{\alpha'ik_2.X(x_2)}\lam\otimes\sigma_1 \nonumber\\
V_{RR}^{(-1)}(z,\bar{z})&=&(P_{-}\fsH_{(n)}M_p)^{\al\be}e^{-\phi(z)/2} S_{\al}(z)e^{i\frac{\alpha'}{2}p\cd X(z)}
e^{-\phi(\bar{z})/2} S_{\be}(\bar{z}) e^{i\frac{\alpha'}{2}p\cd D \cd X(\bar{z})}\otimes \sigma_3\label{vertices}\eeqa
 world-sheet is taken as disk so that all open strings and RR are located on the boundary and the middle of disk accordingly. On-shell conditions are $p^2=0, k_{1}^2=k_{2}^2=1/4$, notations are also followed by
\beqa
P_{-} =\ha (1-\ga^{11}),
\fsH_{(n)} = \frac{a
_n}{n!}H_{\mu_{1}\ldots\mu_{n}}\ga^{\mu_{1}}\ldots
\ga^{\mu_{n}},
(P_{-}\fsH_{(n)})^{\al\be} =
C^{\al\del}(P_{-}\fsH_{(n)})_{\del}{}^{\be}.
\eeqa
where  for type IIA  (type IIB) $n=2,4$,$a_n=i$  ($n=1,3,5$,$a_n=1$) stands. To deal with standard holomorphic functions and based on various change of variables, the doubling trick has been set as follows 
\begin{displaymath}
\tilde{X}^{\mu}(\bar{z}) \rightarrow D^{\mu}_{\nu}X^{\nu}(\bar{z}) \ ,
\spa
\tilde{\psi}^{\mu}(\bar{z}) \rightarrow
D^{\mu}_{\nu}\psi^{\nu}(\bar{z}) \ ,
\spa
\tilde{\phi}(\bar{z}) \rightarrow \phi(\bar{z})\,, \mand
\tilde{S}_{\al}(\bar{z}) \rightarrow M_{\al}{}^{\be}{S}_{\be}(\bar{z})
 \ ,
\non\end{displaymath}

with further ingredients as 
\begin{displaymath}
D = \left( \begin{array}{cc}
-1_{9-p} & 0 \\
0 & 1_{p+1}
\end{array}
\right) \ ,\,\, \mand
M_p = \left\{\begin{array}{cc}\frac{\pm i}{(p+1)!}\ga^{i_{1}}\ga^{i_{2}}\ldots \ga^{i_{p+1}}
\eps_{i_{1}\ldots i_{p+1}}\,\,\,\,{\rm for\, p \,even}\\ \frac{\pm 1}{(p+1)!}\ga^{i_{1}}\ga^{i_{2}}\ldots \ga^{i_{p+1}}\ga_{11}
\eps_{i_{1}\ldots i_{p+1}} \,\,\,\,{\rm for\, p \,odd}\end{array}\right.
\non\end{displaymath}
\vskip .2in

Making use of the above doubling trick, we can now head off and start working with the following two point functions for all the fields of  $X^{\mu},\psi^\mu, \phi$ as follows
\begin{eqnarray}
\lan X^{\mu}(z)X^{\nu}(w)\ran & = & -\frac{\alpha'}{2}\eta^{\mu\nu}\log(z-w) \ , \non \\
\lan \psi^{\mu}(z)\psi^{\nu}(w) \ran & = & -\frac{\alpha'}{2}\eta^{\mu\nu}(z-w)^{-1} \ ,\non \\
\lan\phi(z)\phi(w)\ran & = & -\log(z-w) \ .
\labell{prop2}\end{eqnarray}
Using gauge fixing as   $(x_1,x_2,z,\bar z)=(x,-x,i,-i)$ and taking $u = -\frac{\alpha'}{2}(k_1+k_2)^2$, the ultimate form of the amplitude is \footnote{ $\alpha'=2$ is set. }
\beqa
{\cal A}^{C^{-1}T^{-1}T^{0}} & \sim & 4 k_{2a}\int_{-\infty}^{\infty} dx (2x)^{-2u-1}
(1+x^{2})^{2 u} \Tr(P_{-}\fsH_{(n)}M_p\gamma^{a})\nonumber\eeqa
and the final result can be read  \cite{Kennedy:1999nn}
 \beqa
{\cal A}^{C^{-1}T^{-1}T^{0}} &=&  \frac{i\mu_p}{4} 2\pi \frac{\Gamma(-2u)}{\Gamma(1/2-u)^2}\Tr(P_{-}\fsH_{(n)}M_p\gamma^a)k_{2a}
 \label{yy1}\eeqa


where $ \mu_p $ is RR charge  and the trace 
for $\gamma^{11}$  kept fixed for the following as well   
\beqa
  p>3 , H_n=*H_{10-n} , n\geq 5.
  \nonumber\eeqa

We actually discussed all the proper expansion of S-matrices within detail in  \cite{Hatefi:2012wj}, so that the entire S-matrix elements are recovered in an EFT by sending either $k_i.k_j\rightarrow 0$ or $(k_i+k_j)^2\rightarrow 0$ which means that one indeed is able to regenerate  massless /tachyonic  singularities of  different configurations. We knew that
 a non-zero coupling   $C_{p-1}\wedge F$ exists, and also there is non-vanishing  two tachyons and a gauge field coupling, thus using momentum conservation along the world volume of brane the correct expansion for D-brane-anti D-brane was explored to be \beqa
u=-p^ap_a\rightarrow 0.
\label{CTT2}\eeqa
 as also clarified in \cite{Hatefi:2012cp}. Note that as we have seen in an RR and a tachyon amplitude this constraint for non- BPS branes gets enhanced to   
 \beqa
 p^ap_a\rightarrow -m^2_{T}=\frac{1}{4}\label{CT1}\eeqa
  The expansion around $u\rightarrow 0$ is 
  \beqa
  2\pi\frac{\Ga(-2u)}{\Ga(1/2-u)^2}
 = -\frac{1}{u}+\sum_{m=-1}^{\infty}c_m(u )^{m+1}, 
 c_{-1}&=&4ln(2), c_0=(\frac{\pi^2}{6}-8ln(2)^2),..\nonumber\
 \eeqa
The only u-channel massless gauge field pole of this particular $CTT$ S-matrix can be regenerated by  following  sub-amplitude in an EFT
\beqa
{\cal A}&=&V_a(C_{p-1},A^{(1)})G_{ab}(A)V_b(,A^{(1)},T_1,T_2)+V_a(C_{p-1},A^{(2)})G_{ab}(A)V_b(,A^{(2)},T_1,T_2)\nonumber\eeqa
 where the presence of Chern-Simons coupling on the brane anti-brane system is needed as follows
 \beqa
  i\mu_p (2\pi\alpha')\int_{\Sigma_{p+1}} \epsilon^{a_0...a_{p}}\bigg(\Tr(C_{a_0...a_{p-2}}d_{a_{p-1}}(A_{1a_{p}}-A_{2a_{p}}))\bigg),
    \label{jj12}\eeqa
The off-shell propagator must be $A^{(1)}$ and $A^{(2)}$ to be able to have consistent result with both string and an EFT. The propagator and vertices are accordingly defined by
\beqa
G_{ab}(A) &=&\frac{i\delta_{ab}}{(2\pi\alpha')^2 T_p
\left(k^2\right)}\nonumber\\
V_b(A^{(1)},T_1,T_2)&=&iT_p(2\pi\alpha')(k_1-k_2)_{b}\nonumber\\
V_b(A^{(2)},T_1,T_2)&=&-iT_p(2\pi\alpha')(k_1-k_2)_{b}\nonumber\\
V_a(C_{p-1},A^{(1)})&=&i\mu_p(2\pi\alpha')\frac{1}{(p-1)!}\epsilon_{a_0\cdots a_{p-1}a}C^{a_0\cdots a_{p-2}} k^{a_{p-1}}\nonumber\\
V_a(C_{p-1},A^{(2)})&=&-i\mu_p(2\pi\alpha')\frac{1}{(p-1)!}\epsilon_{a_0\cdots a_{p-1}a}C^{a_0\cdots a_{p-2}}k^{a_{p-1}}\eeqa
Taking into account  $k^a=(k_1+k_2)^{a}=-p^{a}$ as well as replacing  them in EFT amplitude, we gain the field theory amplitude as follows
\beqa
{\cal A}&=&4i\mu_p\frac{1}{p!u}\epsilon^{a_0\cdots a_{p-1}a}H_{a_0\cdots a_{p-1}}k_{2a} \labell{amp444}\eeqa
which is exactly as derived in string theory amplitude \reef{yy1}. In  \cite{Hatefi:2016yhb} we also constructed all order $\alpha'$ corrections of one RR and two tachyons of D-brane anti D-brane system as
\beqa
i\mu_p (2\pi\alpha')^2  C_{(p-1)}\wedge \Tr\left(\sum_{m=-1}^{\infty}c_m(\alpha' (D ^b D_b))^{m+1}  DT \wedge DT^*\right) \labell{highaaw3}\eeqa

\section{  All order  $<V_{C^{-1}} V_{T^{-1}} V_{T^{0}}V_{T^{0}}>$ S-Matrix }
 
All the correlators of this S-matrix can be investigated. We define 
\beqa
s&=&\frac{-\alpha'}{2}(k_1+k_3)^2,\quad t=\frac{-\alpha'}{2}(k_1+k_2)^2,\quad u=\frac{-\alpha'}{2}(k_2+k_3)^2
\nonumber\eeqa 
One finds out the amplitude after the gauge fixing $(x_1,x_2,x_3,z,\bar{z})=(0,1,\infty,z,\bar{z})$  to be 
\beqa
{\cal A}^{C^{-1} T^{-1} T^{0} T^{0}}&\sim&\int \int dzd\bar z (P_{-}\fsH_{(n)}M_p)^{\al\be} \
(-4i k_{2a}k_{3b})x_{45}^{-2(t+s+u+1)}|z|^{2t+2s}|1-z|^{2t+2u}\nonumber\\&&\times
\bigg[(\Gamma^{ba}C^{-1})_{\alpha\beta}+2\eta^{ab}(C^{-1})_{\alpha\beta}(\frac{1-x}{z-\bar z})\bigg]
\nonumber\eeqa

where
$z=x_4=x+iy, \bar z=x_5=x-iy, x_{ij}=x_i-x_j$. Notice that all integrations on upper half plane are carried out on RR position as follows
\beqa
 \int  \int d^2 \!z |1-z|^{a} |z|^{b} (z - \bar{z})^{c}
(z + \bar{z})^{d}\nonumber
\eeqa
and the final result on those moduli spaces has been verified in detail for  $d=0,1$  in \cite{Fotopoulos:2001pt} and for $d=2$  in \cite{Hatefi:2012wj} .
Using some algebraic analysis  the ultimate form of S-matrix is obtained to be
\beqa {\cal A}^{C^{-1}T^{-1} T^{0} T^{0}}&=&{\cal A}_{1}+{\cal A}_{2}\nonumber\\
{\cal A}_{1}&\!\!\!\sim\!\!\!&-4i k_{2a} k_{3b}\Tr(P_{-}\fsH_{(n)}M_p\Gamma^{ba})N_1
\nonumber\\
{\cal A}_{2}&\sim&-4i\Tr(P_{-}\fsH_{(n)}M_p)N_2 
\nonumber\eeqa
where the functions
 $N_1,N_2$ are given as \beqa
N_1&=&(2)^{-2(t+s+u+1)}\pi{\frac{\Gamma(-u)
\Gamma(-s)\Gamma(-t)\Gamma(-t-s-u-\frac{1}{2})}
{\Gamma(-u-t)\Gamma(-t-s)\Gamma(-s-u)}},\label{kk}\\
N_2&=&(2)^{-2(t+s+u)-3}\pi{\frac{\Gamma(-u+\frac{1}{2})
\Gamma(-s+\frac{1}{2})\Gamma(-t+\frac{1}{2})\Gamma(-t-s-u-1)}
{\Gamma(-u-t)\Gamma(-t-s)\Gamma(-s-u)}}
\nonumber\eeqa
One normalizes the amplitude by $3\pi^{-1/2}\beta'\mu_p'/8$ to be able to get consistent result with EFT. As can be explicitly seen, the amplitude is entirely symmetric with respect to interchanging  $s,t,u$ and for $p-1=n$ case and it has an infinite massless gauge field poles in all $s, t, u-$ channels and also for $p+1=n$ case it does have an infinite tachyon poles in $(s+t+u+1)-$channel poles( $s'=s+\frac{1}{2},t'=t+\frac{1}{2},u'=u+\frac{1}{2})$. Given these arguments, the fact that the on-shell condition is
\beqa
s+t+u&=&-p_ap^a-\frac{3}{4}\label{ex1}\eeqa 
and  $p_ap^a\rightarrow 1/4$ for non-BPS D-branes, one obtains uniquely the expansion for $CTTT$ as
\beqa
\frac{1}{3}\bigg(
(u\rightarrow 0,\, s,t\rightarrow -1/2),
 (s\rightarrow 0,\, u,t\rightarrow -1/2),
(t\rightarrow 0,\, s,u\rightarrow -1/2)\bigg)
\label{ctttex}\eeqa


The expansion of $N_1$ around \reef{ex1} is
\beqa
N_1=
\frac{2\pi\sqrt{\pi}}{3}\left(-\frac{1}{u}\sum_{n=-1}^{\infty}b_n(s'+t')^{n+1}
+\sum_{p,n,m=0}^{\infty}c_{p,n,m}u^p\left(s't'\right)^n(s'+t')^m+(u\leftrightarrow t)+(u\leftrightarrow s)\right)\nonumber\eeqa
where some coefficients are 
\beqa
b_{-1}=1,c_{p,0,0}=a_p, a_0=4\ln(2), a_1=\frac{\pi^2}{6}-8\ln(2)^2, c_{0,1,0}=-14\z(3)\eeqa

The string amplitude for this case is given by 
\beqa
\frac{24\beta'\mu_p'}{\sqrt{\pi}(p-1)!u}\sum_{n=-1}^{\infty}b_n(s'+t')^{n+1} \eps^{a_{0}\cdots a_{p}}H_{a_{0}\cdots a_{p-2}}
k_{2a_{p-1}}k_{3a_{p}}\labell{pp}\eeqa

These infinite u-channel gauge field poles can be produced in an EFT by the following sub-amplitude  
\beqa
{\cal A}&=&V_a^{\alpha}(C_{p-2},T_1,A)G_{ab}^{\alpha\beta}(A)V_b^{\beta}(A,T_2,T_3)\labell{amp3}\eeqa
where the vertices are 
\beqa
G^{\alpha\beta}_{ab}(A) &=&\frac{i\delta_{ab}\delta_{\alpha\beta}}{(2\pi\alpha')^2 T_p u}\nonumber\\
V^{\beta}_b(A,T_2,T_3)&=&iT_p(2\pi\alpha')(k_2-k_3)_b \Tr(\lam_2\lam_3\Lambda^{\beta}) \nonumber\\
V^{\alpha}_a(C_{p-2},T_1,A)&=&2\beta'\mu_p'(2\pi\alpha')^2\frac{1}{(p-1)!}\epsilon_{a_1\cdots a_{p}a}H^{a_1\cdots a_{p-1}}k_1^{a_{p}}\sum_{n=-1}^{\infty}b_n(\alpha'k_1\cdot k)^{n+1}\Tr(\lam_1\Lambda^{\alpha})\nonumber\eeqa
  $k$ is off-shell 's gauge field momentum. Finally all contact interactions of this part of S-matrix or all order higher derivative corrections of $C_{p-2}\wedge DT\wedge DT\wedge DT$ can be derived by
 the following coupling
\beqa
&&8\beta'(\pi\alpha'^2)\mu_p'\sum_{p,n,m=0}^{\infty}c_{p,n,m}\left(\frac{\alpha'}{2}\right)^{p}\left(\alpha'\right)^{2n+m} C_{p-2}\wedge \Tr\left(\frac{}{}D^{a_1}\cdots D^{a_{2n}} D^{b_1}\cdots D^{b_{m}}DT\right.\nonumber\\
&&\left.\wedge(D^aD_a)^p  D_{b_1}\cdots D_{b_{m}}(D_{a_1}\cdots D_{a_n}DT\wedge D_{a_{n+1}}\cdots D_{a_{2n}}DT)\frac{}{}\right)\labell{hderv12}\eeqa
with  $\beta'=\frac{1}{\pi}\sqrt{\frac{6\ln(2)}{\alpha'}}$  becomes normalization constant of WZ effective action of non-BPS branes. On the other hand all infinite tachyon channel poles in string side can be explored as  
\beqa
&&\bigg(-1+\frac{1}{3}\sum_{n,m=0}^{\infty}e_{n,m}[(s'+t')^n(t's')^{m+1}+(t',s'\rightarrow t',u')+(t',s'\rightarrow s',u')]\bigg)\nonumber\\&&\times
\frac{24i\beta'\mu_p'}{(p+1)! (t+s+u+1)} \eps^{a_{0}\cdots a_{p}}H_{a_{0}\cdots a_{p}}\labell{pop21}\eeqa
with \beqa
e_{0,0}=-\pi^2/3,e_{1,0}=8\z(3), e_{0,1}=\pi^4/45, e_{1,1}=-32\z(5)+8\z(3)\pi^2/3\nonumber\eeqa

All infinite tachyon poles are produced in an EFT by the following couplings and vertices 
\beqa
{\cal A}&=&V^{\alpha}(C_{p},T)G^{\alpha\beta}(T)V^{\beta}(T,T,T,T)\labell{amp4}\eeqa
 The propagator and vertex  $ V(C_{p},T)$ are
\beqa
G^{\alpha\beta}(T) &=&\frac{i\delta_{\alpha\beta}}{(2\pi\alpha') T_p
\left(-k^2-m^2\right)}\nonumber\\
V^{\alpha}(C_{p},T)&=&2i\beta'\mu_p'(2\pi\alpha')\frac{1}{(p+1)!}\epsilon_{a_0\cdots a_{p}}H^{a_0\cdots a_{p}}\Tr(\Lambda^{\alpha})\labell{Fey55}\nonumber
\eeqa
Following Lagrangian is also needed
\beqa
-T_p\Tr\left((\pi\alpha')m^2T^2+(\pi\alpha')D_aTD^aT-(\pi\alpha')^2F_{ab}F^{ba}+T^4\right)\labell{4T1}\eeqa 

Using   \reef{4T1} and the fact that off-shell tachyon in propagator is abelian, the vertex $V^{\beta}(T,T,T,T)$ is constructed to be  $V^{\beta}(T,T,T,T)=-12iT_p \Tr(\lam_1\lam_2\lam_3\Lambda^{\beta})$. Considering it in \reef{amp4}, one reveals the tachyon pole of string amplitude in an EFT to be \beqa
24i\beta'\mu_p'\frac{1}{(p+1)! (s+t+u+1)}\epsilon_{a_0\cdots a_{p}}H^{a_0\cdots a_{p}}\labell{Fey133}\nonumber
\eeqa
In the next section we would like to perform an RR and four tachyon couplings on 
 D-brane-anti-D-brane's world volume which is the generalization of Veneziano amplitude of four open string tachyons
 \cite{Veneziano:1968yb}. We observe that all infinite tachyon poles are reconstructed by 
 $ (2\pi\alpha')^2 \mu_p \beta^2 \int C_{p-1}\wedge DT\wedge DT$ and all order extensions of four tachyon couplings.
 
   \section{ $<V_{C^{-1}(z,\bar z)}V_{T^{-1} (x_1)}V_{T^{0}(x_2)}V_{T^{0}(x_3)}V_{T^{0}(x_4)}>$  Amplitude}

In this section , given all symmetries that we discussed in the last sections and the fact that tachyonic expansion has ben checked for all lower point functions of string amplitudes on the world volume of D-brane-Anti-D-brane systems such as  $CTT, CTT\phi,CTTA$ etc, we intend to show that all the infinite singularities of an RR and four tachyons on the  world volume of D-brane-Anti-D-brane  can be constructed out, although the precise and algebraic forms of integrals are unknown.Various techniques have been demonstrated, fixing the position of open strings at $x_1=0, 0\leq x_2\leq 1 , x_3=1, x_4=\infty$ and using 6 independent Mandelstam variables as $s=-(\frac{1}{2}+2k_1.k_3), t=-(\frac{1}{2}+2k_1.k_2), v=-(\frac{1}{2}+2k_1.k_4), u=-(\frac{1}{2}+2k_2.k_3), r=-(\frac{1}{2}+2k_2.k_4), w=-(\frac{1}{2}+2k_3.k_4)$ the closed form of amplitude is   
 
\beqa 
&&8 k_{2a}k_{3b}k_{4c} 2^{-1/2}(P_{-}\fsH_{(n)}M_p)^{\alpha\beta}\int_{0}^{1} dx_2 x_2^{-2t-1} (1-x_2)^{-2u-1}\int \int d^2z |1-z|^{2s+2u+2w+1} |z|^{2t+2s+2v+1}
\nonumber\\&&\times(z - \bar{z})^{-2(t+s+u+v+r+w+2)}  |x_2-z|^{2t+2u+2r+1}
\bigg[(\Gamma^{cba}C^{-1})_{\alpha\beta}+(z - \bar{z})^{-1}\bigg(2\eta^{ab}(\gamma^{c}C^{-1})_{\alpha\beta}\nonumber\\&&\times(1-x_2)^{-1}(x_2-xx_2-x+|z|^{2})-2\eta^{ac}(\gamma^{b}C^{-1})_{\alpha\beta}(x_2-x)+2\eta^{bc}(\gamma^{a}C^{-1})_{\alpha\beta} (1-x)\bigg) \bigg] \label{eerr33}\eeqa

Hence $CTTTT$ S-matrix makes sense for $p-2=n, p=n$ cases accordingly. If we use the particular limit $4k_2.p\rightarrow 1$ that we called soft limit\footnote{ As can be seen it is consistent with EFT couplings and produces all massless singularities of the S-matrix.} then one reveals that the algebraic solutions for all the integrals on upper half plane can be entirely achieved, as explained in \cite{Hatefi:2012wj}. In this particular limit, the ultimate form of the amplitude for $p-2=n$ case will be given by
\beqa
{\cal A}_{1}&\sim& 2^{-2(t+s+u+v+r+w)} \pi  2^{-1/2}k_{2a} k_{3b}k_{4c}\frac{8}{(p-2)!}\epsilon^{a_0...a_{p-3}cba} H_{a_0...a_{p-3}} M_1,\label{kk88}\\
M_1&=&{\frac{\Gamma(-2u)\Gamma(-2t)
\Gamma(r-s)\Gamma(-t-v-r-\frac{1}{2})\Gamma(-u-r-w-\frac{1}{2})\Gamma(-t-s-u-v-r-w-\frac{3}{2})}
{\Gamma(-2t-2u)\Gamma(-t-s-v-\frac{1}{2})\Gamma(-u-s-w-\frac{1}{2})\Gamma(-t-u-v-w-2r-1)}}\nonumber\eeqa
The second part of the amplitude makes sense for $p=n$ case and after simplifications we obtain it as follows
\beqa
{\cal A}_{2}&\sim& 2^{-2(t+s+u+v+r+w)} \pi  2^{-1/2}\frac{2}{p!}\epsilon^{a_0...a_{p-1}a} H_{a_0...a_{p-1}} \bigg\{-2k_{2a}(w+\frac{1}{2})(r-s-\frac{1}{2})M_3\nonumber\\&&+4k_{3a}(r+\frac{1}{2})M_2
\bigg(t(1+t+r+v)-u(1+r+u+w)\bigg)
 \nonumber\\&&+k_{4a}M_3\bigg((1+2u)(1+r+u+w)+t+2t(1+u+w+s)\bigg) \bigg\},\nonumber\\
M_2&=&{\frac{\Gamma(-2u)\Gamma(-2t)
\Gamma(r-s+\frac{1}{2})\Gamma(-t-v-r-1)\Gamma(-u-r-w-1)\Gamma(-t-s-u-v-r-w-2)}
{\Gamma(1-2t-2u)\Gamma(-t-s-v-\frac{1}{2})\Gamma(-u-s-w-\frac{1}{2})\Gamma(-t-u-v-w-2r-1)}}\nonumber\\
M_3&=&{\frac{\Gamma(-2u)\Gamma(-2t)
\Gamma(r-s-\frac{1}{2})\Gamma(-t-v-r)\Gamma(-u-r-w-1)\Gamma(-t-s-u-v-r-w-2)}
{\Gamma(-2t-2u)\Gamma(-t-s-v-\frac{1}{2})\Gamma(-u-s-w-\frac{1}{2})\Gamma(-t-u-v-w-2r-1)}}\nonumber
\eeqa

Applying momentum conservation along the brane we derive 
\beqa
 s+t+u+v+r+w=-p^a p_a-2\label{rr33}\eeqa
  Given \reef{rr33}, the EFT methods that imposed to have an infinite massless either u or t- channel gauge field poles, also highly symmetries \cite{Schwarz:2013wra} of this amplitude ( it should be symmetric under exchanging Mandelstam variables $u,t$ and other variables),  $k_i.k_j\rightarrow 0$  and the fact that for tachyonic strings $p^ap_a \rightarrow \frac{1}{4}$ we obtain the following expansion for this particular soft limit  of the scattering amplitude  as follows

 \begin{displaymath}
\frac{1}{4}\bigg( v,w\rightarrow -\frac{1}{2}\bigg)\left\{\begin{array}{cccc}
u\rightarrow 0, s\rightarrow -\frac{1}{4}, (t,r,\rightarrow -\frac{1}{2})\\
\\

u\rightarrow 0, r\rightarrow -\frac{1}{4}, (t,s,\rightarrow -\frac{1}{2})\label{ufexp}\\
\\

t\rightarrow 0, s\rightarrow -\frac{1}{4}, (u,r,\rightarrow -\frac{1}{2})\\
\\

t\rightarrow 0, r\rightarrow -\frac{1}{4}, (u,s,\rightarrow -\frac{1}{2})\end{array}\right.
\end{displaymath}

Given the so called selection rules for non-BPS couplings of string theory \cite{Hatefi:2013yxa} and also the fact that all kinetic terms have already been fixed in DBI action, one comes to know that for $p-2=n$ case we have just an infinite single massless gauge field poles in both u and t- channels. The expansion of $M_1$ around the first above  expansion is
 
 \beqa
 && 2^{1/2}\Gamma(3/4)^{2}\bigg(\frac {2}{u}-8\bigg(\frac{\ln2}{2}+\frac{\pi}{4}-\frac{1}{2}\bigg)-\frac{8}{u}\bigg((r+t+v+\frac{s+w}{2})\ln2\nonumber\\&&
 +\frac{s+w}{4}\pi -2r+s-\frac{t+v+w}{2}\bigg)+...\bigg)\label{plm}\eeqa
 
 As it is clear from \reef{kk88} and $M_1$ for $p-2=n$  case the amplitude has an infinite u,t-channel massless gauge field poles and due to symmetries we just produce u-channel poles as follows. 
 \beqa
V^{a}(C_{p-3},A^{(1)},T_1,T_4)G^{ab}(A)V^{b}(A^{(1)},T_2,T_3)+V^{a}(C_{p-3},A^{(2)},T_1,T_4)G^{ab}(A)V^{b}(A^{(2)},T_2,T_3)\nonumber
\eeqa
\beqa
G^{ab}(A) &=&\frac{i\delta^{ab}}{(2\pi\alpha')^2 T_p u}\nonumber\\
V^{b}(A^{(1)},T_2,T_3)&=&i T_p (2\pi\alpha') (k_2-k_3)^{b}\nonumber\\
V^{b}(A^{(2)},T_2,T_3)&=&-i T_p (2\pi\alpha') (k_2-k_3)^{b}\nonumber\\
V^{a}(C_{p-3},A^{(1)},T_1,T_4)&=&i\mu'_p\beta (2\pi\alpha')^3\frac{1}{(p-2)!}\epsilon^{a_0\cdots a_{p-1}a}H_{a_0\cdots a_{p-3}} k_{4a_{p-2}}k_{a_{p-1}}\nonumber\\
V^{a}(C_{p-3},A^{(2)},T_1,T_4)&=&-i\mu'_p\beta (2\pi\alpha')^3\frac{1}{(p-2)!}\epsilon^{a_0\cdots a_{p-1}a }H_{a_0\cdots a_{p-3}} k_{4a_{p-2}}k_{a_{p-1}} 
\labell{vvxz551}
\eeqa
 where $k_b= (k_2+k_3)_b$ is the momentum of off-shell gauge field. Note that $V^{a}(C_{p-3},A,T_1,T_4)$  was derived from coupling 
 \beqa
 \beta\mu_p (2\pi\alpha')^{3}\int_{\Sigma_{p+1}}\Tr (C_{p-3}\wedge F\wedge DT_1\wedge DT_4)\label{bb12}\eeqa  
 
 If we normalize the amplitude by $\frac{2^{3/2}\beta\mu_p (2\pi)^{2}}{\Gamma(3/4)^{2}}$, then one observes that the first u-channel gauge field pole is reconstructed in an EFT by \reef{vvxz551} to be 
 
 \beqa
 i\mu_p\beta (2\pi\alpha')^2\frac{4}{(p-2)!u}\epsilon^{a_0\cdots a_{p-3}cba }H_{a_0\cdots a_{p-3}} k_{2a} k_{3b}k_{4c} \nonumber\eeqa
 The vertex of $V^{b}(A^{(1)},T_2,T_3)$ and single pole are fixed and have no corrections so to be able to generate all other massless u-channel poles one needs to apply higher derivative corrections to above WZ coupling \reef{bb12}, however, given the entire explanations of the previous sections  we are no longer interested in doing so. We would rather carry out the rest of the analysis of the amplitude which has something to the extensions of four tachyon couplings.
 \vskip.1in

  On the other hand, having set the EFT arguments, one would become aware of the fact that for $p=n$ case, the amplitude has an infinite  $(u+w+r+1)$  channel (and due to symmetries  $(t+s+u+1),(s+v+w+1),(t+v+r+1)
  $ channel as well) tachyon singularity structures and a double pole that can be shown later on.
 
 \vskip.1in
 
 As it is clear from ${\cal A}_{2}$, it has an infinite $(u+w+r+1)$ channel tachyon poles that can be explored in an EFT within the following sub-amplitude as follows
 \beqa
{\cal A}&=&V^{\alpha}(C_{p-1},T,T_1)G^{\alpha\beta}(T)V^{\beta}(T,T_2,T_3,T_4)\nonumber\\
G^{\alpha\beta}(T) &=&\frac{i\delta_{\alpha\beta}}{(2\pi\alpha') T_p
(u+w+r+1)}\nonumber\\
V^{\alpha}(C_{p-1},T,T_1)&=&\mu_p\beta^2 (2\pi\alpha')^2\frac{1}{(p-1)!}\epsilon^{a_0\cdots a_{p}}H_{a_0\cdots a_{p-1}} k_{4a_{p}}\Tr(\lambda_1\Lambda^{\alpha})\labell{4482vv}
\eeqa

To produce an infinite tachyon poles, one needs to employ the  higher derivative corrections to four tachyon  couplings as follows
    
\beqa
T_p(\alpha')^{2+n+m}\sum_{m,n=0}^{\infty}(\cL_{1}^{nm}+\cL_{2}^{nm}+\cL_{3}^{nm}+\cL_{4}^{nm}+\cL_{5}^{nm})\labell{lagrang11}\eeqa
where
  \beqa
\cL_{1}^{nm}&\!\!\!\!\!\!=\!\!\!\!\!\!&m^4\Tr\left(\frac{}{}a_{n,m} D_{nm}[\cT\cT\cT\cT]- b_{n,m} D'_{nm}[\cT\cT\cT\cT]+h.c.\frac{}{}\right)\nonumber\\
\cL_{2}^{nm}&\!\!\!\!\!\!=\!\!\!\!\!\!&m^2\Tr\left(\frac{}{}a_{n,m}[ D_{nm}(\cT \cT D^{\alpha}\cT D_{\alpha}\cT)+D_{nm}( D^{\alpha}\cT D_{\alpha}\cT\cT \cT)]\right.\nonumber\\
&&\left.-b_{n,m}[D'_{nm}(\cT D^{\alpha}\cT  \cT  D_{\alpha}\cT)+D'_{nm}( D^{\alpha}\cT  \cT D_{\alpha}\cT \cT)] +h.c.\frac{}{}\right)\nonumber\\
\cL_{3}^{nm}&\!\!\!\!\!\!=\!\!\!\!\!\!&-\Tr\left(\frac{}{}a_{n,m}D_{nm}[D_{\alpha}\cT D_{\beta}\cT D^{\beta}\cT D^{\alpha}\cT]-b_{n,m}D'_{nm}[D_{\alpha}\cT D^{\beta}\cT D_{\beta}\cT  D^{\alpha}\cT]+h.c.\frac{}{}\right)\nonumber\\
\cL_{4}^{nm}&\!\!\!\!\!\!=\!\!\!\!\!\!&-\Tr\left(\frac{}{}a_{n,m}D_{nm}[D_{\alpha}\cT D_{\beta}\cT D^{\alpha}\cT D^{\beta}\cT] -b_{n,m} D'_{nm}[D_{\beta}\cT D^{\beta}\cT  D_{\alpha}\cT D^{\alpha}\cT]+h.c. \frac{}{}\right)\nonumber\\
\cL_{5}^{nm}&\!\!\!\!\!\!=\!\!\!\!\!\!&\Tr\left(\frac{}{}a_{n,m} D_{nm}[D_{\alpha}\cT D^{\alpha}\cT D_{\beta}\cT D^{\beta} \cT] -b_{n,m} D'_{nm}[D_{\alpha}\cT D_{\beta}\cT D^{\alpha}\cT  D^{\beta}\cT]+h.c. \frac{}{}\right)\nonumber\eeqa

with following definition for $D_{nm}$ and $D'_{nm}$ derivative operators as
\beqa
D_{nm}(EFGH)&\equiv&D_{b_1}\cdots D_{b_m}D_{a_1}\cdots D_{a_n}E  F D^{a_1}\cdots D^{a_n}GD^{b_1}\cdots D^{b_m}H\nonumber\\
D'_{nm}(EFGH)&\equiv&D_{b_1}\cdots D_{b_m}D_{a_1}\cdots D_{a_n}E   D^{a_1}\cdots D^{a_n}F G D^{b_1}\cdots D^{b_m}H\nonumber\eeqa

All order four tachyon extensions of  $V^{\beta}(T,T_2,T_3,T_4)$ might be explored from the higher derivative couplings \reef{lagrang11} as below
\beqa
&&2iT_p
(\alpha')^{n+m}(a_{n,m}-b_{n,m})\Tr(\lam_2\lam_3\lam_4\Lambda^{\beta})\bigg[u'r'\bigg((k_1\inn k)^m(k_2\inn k_4)^n+(k_1\inn k)^n(k_2\inn k_4)^m\nonumber\\&&+(k_2\inn k)^m(k_2\inn k_3)^n+
(k_2\inn k)^n(k_2\inn k_3)^m+(k_1\inn k)^m(k_2\inn k)^n+(k_1\inn k)^n(k_2\inn k)^m\nonumber\\&&+(k_2\inn k_3)^m(k_2\inn k_4)^n+
(k_2\inn k_3)^n(k_2\inn k_4)^m\bigg)+(u,r\rightarrow u,w)+(u,r\rightarrow r,w)\bigg]\nonumber\eeqa
 $k^a$ is off-shell tachyon 's momentum.  Taking into account this vertex in \reef{4482vv}, one gains all tachyon poles in an EFT as follows
\beqa
32i\beta^2\mu_p \eps^{a_{0}\cdots a_{p}}H_{a_{0}\cdots a_{p-1}}k_{4a_{p}} \frac{1}{(p-1)!(u+r+w+1)}
\sum_{n,m=0}^{\infty}(a_{n,m}-b_{n,m})\nonumber\\
\bigg[u'r'(u'^mr'^n+u'^n r'^m)+u'w'(u'^mw'^n+u'^n w'^m)+r'w'(r'^mw'^n+r'^n w'^m)\bigg]\nonumber\eeqa
One may re-write down the above amplitude as follows
\beqa
&&\sum_{n,m=0}^{\infty}e_{n,m} \bigg[(u'+r')^n(u'r')^{m+1}+(u'+w')^n(u'w')^{m+1}+(r'+w')^n(r'w')^{m+1}\bigg]\nonumber\\&&\times
16i\beta^2\mu_p   
\frac{ \eps^{a_{0}\cdots a_{p}}H_{a_{0}\cdots a_{p-1}}k_{4a_{p}}}{(p-1)!(u+r+w+1)}\nonumber\eeqa
These  tachyon poles are exactly the same singularities that appeared in second part of string amplitude. 

\vskip.1in

EFT imposes to us that the amplitude does have just a double pole for $C_{p-1}$ case as
\beqa
 V(C_{p-1},T_1,T_2)G(T_2)V_{a}(T_2,T_4,A)G_{ab}(A)V_b(A,T_2,T_3)\labell{amp777xz}\eeqa
and the off-shell gauge field $A$ needs to be both $A^{(1)}$ and $A^{(2)}$ with the following vertices derived  
\beqa
V(C_{p-1},T_1,T_2)&=&\beta^2\mu_p(2\pi\alpha')^2\frac{1}{p!} \eps^{a_{0}\cdots a_{p-1}c}H_{a_{0}\cdots a_{p-1}}k_{2c}\nonumber\\
V_{a}(T_2,T_4,A)&=&T_p(2\pi\alpha')(k_{2a}+k_a)\nonumber\\
G_{ab}(A)&=&\frac{i \delta^{ab}}{(2\pi\alpha')^2T_p u}\\
V_b(A,T_2,T_3)&=&T_p(2\pi\alpha')(k_{2}-k_{3})_{b}\nonumber\\
G(T_2)&=&\frac{i}{(2\pi\alpha')T_p(u+r+w+1)}\labell{ver2}\eeqa
with $k$ is  off-shell gauge field's momentum. Inserting above vertices in \reef{amp777xz} we are then able to reconstruct the double pole in an EFT as follows 
\beqa
\beta^2 (2\pi\alpha')\mu_p\left(\frac{-1}{u(u+r+w+1)}\right)\frac{1}{2p!} \eps^{a_{0}\cdots a_{p-1}c}H_{a_{0}\cdots a_{p-1}}(p+k_1)_{c} (r-w)\labell{amp51}\eeqa

It is worth to pointing out that the appearance of $F^{(1)}\cdot F^{(2)}$ and also   $D\phi^{i(1)}\cdot D\phi_{i(2)}$ couplings for D-brane-Anti D-brane DBI action have been confirmed  by direct consistent string amplitudes of \cite{Garousi:2007fk} and \cite{Hatefi:2016yhb} accordingly. More crucially, we got to know that $C_{p-1}\wedge F$ does not receive higher derivative correction of WZ couplings. The other important result that we found is that
as the kinetic terms donot  get any corrections hence all non-leading tachyonic singularities provide some information not only about the structure of higher derivative corrections to some couplings such as $C_{p-1}\wedge DT\wedge DT$ but also could fix their coefficients on the world volume of brane-anti brane system as well. We could further move on and talk about all the contact terms of the string amplitude, for instance a new coupling such as  $C_{p-1}\wedge F TT^*$ can also be confirmed. We hope to further analyze the higher derivative corrections to those new couplings in near future and to be able to eventually fix all the higher derivative corrections to brane anti-brane DBI as well as WZ effective actions.

\vskip.2in

Remarks are in order. As we discussed these non-BPS couplings are worked out in the presence of the constraint $p_ap^a\rightarrow 1/4$, hence these couplings cannot be compared with BSFT couplings, although the tachyon potential is the same as the one that appeared in BSFT ($V({T})=e^{\pi\alpha'm^2{ T}^2}$ ~\cite{Kutasov:2000aq}) \ie
 \beqa
V( T^iT^i)&=&1+\pi\alpha'm^2{ T^iT^i}+
\frac{1}{2}(\pi\alpha'm^2{ T^iT^i})^2+\cdots
\non\eeqa  where 
$m_{T}^2=-1/(2\alpha')$. However, note that there is a symmetrized trace on the $\sigma$ factors in DBI action, carrying out the symmetrized trace, one finds
\beqa
\frac{1}{2}\STr\left(V({ T^iT^i})\sqrt{1+[T^i,T^j][T^j,T^i]}\right)&=&\left(1-\frac{\pi}{2}T^2+\frac{\pi^2}{24}T^4+\cdots\right)\left(1+T^4+\cdots\right)\nonumber\eeqa
The tachyon does condensate at $T\rightarrow \infty$ and hence, the tachyon potential goes to zero at that point as well.  

\section*{Acknowledgments}

The author would like to thank L. Alvarez-Gaume, B. Jurco, O. Lechtenfeld, J.Kunz, P.H. Damgaard,  N. E Bjerrum-Bohr, N. Arkani-Hamed, D. Jackson, C. Bachas, D. McGady, H. Steinacker, P. Schupp, J. Polchinski, J. Schwarz, A. Sen, W. Siegel, G. Veneziano and E. Witten for  fruitful discussions.  Some parts of this work are taken place at QMUL in London, CERN, ICTP, Charles University in math dep,  IHES both physics and math deps, Leibniz University Hannover, Oldenburg University and Niels Bohr Institute in Copenhagen. He would like to thank those institutes for warm hospitality.



\begin{thebibliography}{2007}



\bibitem{Gutperle:2002ai}
  M.~Gutperle and A.~Strominger,
  JHEP {\bf 0204}, 018 (2002)
  [arXiv:hep-th/0202210];
  A.~Sen,
  JHEP {\bf 0204}, 048 (2002)
  [arXiv:hep-th/0203211];
  A.~Sen,
  JHEP {\bf 0210}, 003 (2002)
  [arXiv:hep-th/0207105];
  A.~Strominger,
  arXiv:hep-th/0209090;
  F.~Larsen, A.~Naqvi and S.~Terashima,
  JHEP {\bf 0302}, 039 (2003)
  [arXiv:hep-th/0212248];
  M.~Gutperle and A.~Strominger,
  Phys.\ Rev.\  D {\bf 67}, 126002 (2003)
  [arXiv:hep-th/0301038];

\bibitem{Lambert:2003zr}
  N.~D.~Lambert, H.~Liu and J.~M.~Maldacena,
  JHEP {\bf 0703}, 014 (2007)
  [arXiv:hep-th/0303139].
\bibitem{Sen:2004nf}
  A.~Sen,
   ``Tachyon dynamics in open string theory,''
  Int.\ J.\ Mod.\ Phys.\ A {\bf 20}, 5513 (2005)
  [hep-th/0410103];
  A.~Sen,
  hep-th/9904207.
%



  
\bibitem{Sakai:2004cn} 
  T.~Sakai and S.~Sugimoto,
  Prog.\ Theor.\ Phys.\  {\bf 113}, 843 (2005)
  [hep-th/0412141];
  T.~Sakai and S.~Sugimoto,
  Prog.\ Theor.\ Phys.\  {\bf 114}, 1083 (2005)
  [hep-th/0507073].
  T.~Okuda and S.~Sugimoto,
  Nucl.\ Phys.\  B {\bf 647}, 101 (2002)
  [arXiv:hep-th/0208196];
 
\bibitem{Casero:2007ae}
  R.~Casero, E.~Kiritsis and A.~Paredes,
  arXiv:hep-th/0702155.
  A.~Dhar and P.~Nag,
  arXiv:0708.3233 [hep-th].







  
  

\bibitem{Sen:1999md}
  A.~Sen,
  JHEP {\bf 9910}, 008 (1999)
  [arXiv:hep-th/9909062].
\bibitem{Bergshoeff:2000dq}
  E.~A.~Bergshoeff, M.~de Roo, T.~C.~de Wit, E.~Eyras and S.~Panda,
  JHEP {\bf 0005}, 009 (2000)
  [arXiv:hep-th/0003221].




\bibitem{Sen:2002in}
  A.~Sen,
  ``Tachyon matter,''
  JHEP {\bf 0207}, 065 (2002)
  [arXiv:hep-th/0203265];
  A.~Sen,
  Mod.\ Phys.\ Lett.\  A {\bf 17}, 1797 (2002)
  [arXiv:hep-th/0204143];



\bibitem{Garousi:2008ge}
  M.~R.~Garousi and E.~Hatefi,
  JHEP {\bf 0903}, 008 (2009)
  [arXiv:0812.4216 [hep-th]].
\bibitem{Sen:1998sm} 
  A.~Sen,
  JHEP {\bf 9808}, 012 (1998)
  [hep-th/9805170].
 

  
  
  
  
  
  
\bibitem{Michel:2014lva} 
  B.~Michel, E.~Mintun, J.~Polchinski, A.~Puhm and P.~Saad,
  JHEP {\bf 1509}, 021 (2015)
  [arXiv:1412.5702 [hep-th]].












\bibitem{Bergman:1998xv}
  O.~Bergman and M.~R.~Gaberdiel,
  Phys.\ Lett.\ B {\bf 441}, 133 (1998)
  [hep-th/9806155]
  ;
  A.~Sen,
  JHEP {\bf 9809}, 023 (1998)
  [hep-th/9808141];
  M.~Frau, L.~Gallot, A.~Lerda and P.~Strigazzi,
  Nucl.\ Phys.\ B {\bf 564}, 60 (2000)
  [hep-th/9903123]
  ;
  E.~Dudas, J.~Mourad and A.~Sagnotti,
  Nucl.\ Phys.\ B {\bf 620}, 109 (2002)
  [hep-th/0107081]
  ;
  E.~Eyras and S.~Panda,
  Nucl.\ Phys.\ B {\bf 584} (2000) 251
  [hep-th/0003033]
  ;
  E.~Eyras and S.~Panda,
  JHEP {\bf 0105}, 056 (2001)
  [hep-th/0009224];
  E.~Hatefi,
  Phys.\ Lett.\ B {\bf 760} (2016) 509
  [arXiv:1511.04971 [hep-th]].
  A.~Lerda and R.~Russo,
  Int.\ J.\ Mod.\ Phys.\ A {\bf 15} (2000) 771
  [hep-th/9905006].
  V.~Calo, G.~Tallarita and S.~Thomas,
  JHEP {\bf 1011}, 063 (2010)
  [arXiv:1003.6063 [hep-th]].
 




\bibitem{Dvali:1998pa}
  G.~R.~Dvali and S.~H.~H.~Tye,
  ``Brane inflation,''
  Phys.\ Lett.\ B {\bf 450} (1999) 72
  [hep-ph/9812483].
;
  D.~Choudhury, etal
  JCAP {\bf 0307}, 009 (2003)
  [hep-th/0305104].
;
  S.~Kachru, etal   ``Towards inflation in string theory,''
  JCAP {\bf 0310}, 013 (2003)
  [hep-th/0308055].
  
  
  
  
  




  

\bibitem{Bjerrum-Bohr:2014qwa} 
  N.~E.~J.~Bjerrum-Bohr, P.~H.~Damgaard, P.~Tourkine and P.~Vanhove,
  Phys.\ Rev.\ D {\bf 90}, no. 10, 106002 (2014)
  [arXiv:1403.4553 [hep-th]].
  E.~Hatefi,
  Eur.\ Phys.\ J.\ C {\bf 74}, no. 10, 3116 (2014)
  [arXiv:1403.7167 [hep-th]];
  L.~A.~Barreiro and R.~Medina,
  Nucl.\ Phys.\ B {\bf 886}, 870 (2014)
  [arXiv:1310.5942 [hep-th]].
  E.~Hatefi,
  Eur.\ Phys.\ J.\ C {\bf 74} (2014) 2949
  [arXiv:1403.1238 [hep-th]];
  O.~Chandia and R.~Medina,
  JHEP {\bf 0311}, 003 (2003)
  [hep-th/0310015];
  E.~Hatefi,
  arXiv:1703.09996 [hep-th].
  E.~Hatefi,
  JHEP {\bf 1703} (2017) 019
  [arXiv:1609.01385 [hep-th]];
  E.~Hatefi,
  Phys.\ Lett.\ B {\bf 766} (2017) 153
  [arXiv:1611.00787 [hep-th]];
  E.~Hatefi,
  Eur.\ Phys.\ J.\ C {\bf 74} (2014) 8,  3003
  [arXiv:1310.8308 [hep-th]].
  E.~Hatefi,
  Phys.\ Lett.\ B {\bf 761}, 287 (2016)
  [arXiv:1604.03514 [hep-th]].

\bibitem{Hatefi:2012zh}
  E.~Hatefi,
  JHEP {\bf 1304}, 070 (2013)
  [arXiv:1211.2413 [hep-th]].














\bibitem{Hatefi:2012ve}
  E.~Hatefi and I.~Y.~Park,
  Phys.\ Rev.\ D {\bf 85} (2012) 125039
  [arXiv:1203.5553 [hep-th]].
;
  E.~Hatefi,
  Nucl.\ Phys.\ B {\bf 880}, 1 (2014)
  [arXiv:1302.5024 [hep-th]].


\bibitem{Hatefi:2012rx}
  E.~Hatefi and I.~Y.~Park,
  Nucl.\ Phys.\ B {\bf 864} (2012) 640
  [arXiv:1205.5079 [hep-th]];
  E.~Hatefi,
  JHEP {\bf 1307}, 002 (2013)
  [arXiv:1304.3711 [hep-th]].
  

\bibitem{Hatefi:2012bp}
  E.~Hatefi, A.~J.~Nurmagambetov and I.~Y.~Park,
  JHEP {\bf 1304}, 170 (2013)
  [arXiv:1210.3825 [hep-th]];
  E.~Hatefi, A.~J.~Nurmagambetov and I.~Y.~Park,
  Nucl.\ Phys.\ B {\bf 866}, 58 (2013)
  [arXiv:1204.2711 [hep-th]].


 


  
  
  
\bibitem{Polchinski:2015bea} 
  J.~Polchinski, ``Brane/antibrane dynamics and KKLT stability,''
  arXiv:1509.05710 [hep-th];
  S.~de Alwis, R.~Gupta, E.~Hatefi and F.~Quevedo,
  JHEP {\bf 1311}, 179 (2013)
  [arXiv:1308.1222 [hep-th].
\bibitem{Polchinski:1995mt}
  J.~Polchinski,
  Phys.\ Rev.\ Lett.\  {\bf 75}, 4724 (1995)
  [hep-th/9510017].
\bibitem{Witten:1995im}
  E.~Witten,
  Nucl.\ Phys.\ B {\bf 460}, 335 (1996)
  [hep-th/9510135].




\bibitem{Hatefi:2010ik}
  E.~Hatefi,
  JHEP {\bf 1005}, 080 (2010)
  [arXiv:1003.0314 [hep-th]].
\bibitem{Myers:1999ps}
  R.~C.~Myers,``Dielectric branes,''
  JHEP {\bf 9912} (1999) 022
  [hep-th/9910053].





\bibitem{Kennedy:1999nn} 
  C.~Kennedy and A.~Wilkins,
  Phys.\ Lett.\ B {\bf 464}, 206 (1999)
  [hep-th/9905195].


\bibitem{Hatefi:2013yxa} 
  E.~Hatefi,
  ``Selection Rules and RR Couplings on Non-BPS Branes,''
  JHEP {\bf 1311}, 204 (2013)
  [arXiv:1307.3520].


\bibitem{Garousi:2007fk}
  M.~R.~Garousi and E.~Hatefi,
  Nucl.\ Phys.\ B {\bf 800} (2008) 502
  [arXiv:0710.5875 [hep-th]].
  
  
  


\bibitem{Sen:2003tm}
  A.~Sen,
  Phys.\ Rev.\  D {\bf 68}, 066008 (2003)
  [arXiv:hep-th/0303057].
  
  
  
  
  
\bibitem{Hatefi:2012cp} 
  E.~Hatefi,
  ``On D-brane anti D-brane effective actions and their corrections to all orders in alpha-prime,''
  JCAP {\bf 1309}, 011 (2013)
  [arXiv:1211.5538,[hep-th]].





\bibitem{Kutasov:2000aq}
  D.~Kutasov, M.~Marino and G.~W.~Moore,
  arXiv:hep-th/0010108.

 


\bibitem{Douglas:1995bn}
  M.~R.~Douglas,
  arXiv:hep-th/9512077;
  M.~Li,
  Nucl.\ Phys.\  B {\bf 460}, 351 (1996)
  [arXiv:hep-th/9510161];
  M.~B.~Green, J.~A.~Harvey and G.~W.~Moore,
  Class.\ Quant.\ Grav.\  {\bf 14}, 47 (1997)
  [arXiv:hep-th/9605033].

\bibitem{Kraus:2000nj}
  P.~Kraus and F.~Larsen,
  Phys.\ Rev.\  D {\bf 63}, 106004 (2001)
  [arXiv:hep-th/0012198];                                       
  T.~Takayanagi, S.~Terashima and T.~Uesugi,
  JHEP {\bf 0103}, 019 (2001)
  [arXiv:hep-th/0012210].
\bibitem{quil}D. Quillen, 
``{\em Superconnections and the Chern character},''
 Topology~{\bf 24} 89--95, (1995).
\bibitem{berl}
N. Berline, E. Getzler and M. Vergne,
`` {\em Heat Kernels and Dirac operators},''
 Springer-Verlag (1991).
\bibitem{Roepstorff:1998vh}
  G.~Roepstorff,
  J.\ Math.\ Phys.\  {\bf 40}, 2698 (1999)
  [arXiv:hep-th/9801040].









\bibitem{Hatefi:2015gwa}
  E.~Hatefi,
  Eur.\ Phys.\ J.\ C {\bf 75} (2015) no.11,  517
  [arXiv:1502.06536 [hep-th]].
 
 
 
\bibitem{Friedan:1985ge}
  D.~Friedan, E.~J.~Martinec and S.~H.~Shenker,
  Nucl.\ Phys.\ B {\bf 271} (1986) 93.
  
  
  
  
  
\bibitem{Hatefi:2012wj}
  E.~Hatefi,
  Phys.\ Rev.\ D {\bf 86} (2012) 046003
  [arXiv:1203.1329 [hep-th]].
  
  
  
\bibitem{Hatefi:2016yhb}
  E.~Hatefi,
   ``On D-brane-Anti D-brane Effective actions and their all order Bulk Singularity Structures,''
  JCAP {\bf 1604} (2016) no.04,  055
  [arXiv:1601.06667 [hep-th]].
  
  
\bibitem{Fotopoulos:2001pt} 
  A.~Fotopoulos,
  JHEP {\bf 0109}, 005 (2001)
  [hep-th/0104146].

  
  
 

\bibitem{Veneziano:1968yb}
  G.~Veneziano,
  Nuovo Cim.\ A {\bf 57} (1968) 190.

 
 
\bibitem{Schwarz:2013wra}
  J.~H.~Schwarz,
  ``Highly Effective Actions,''
  JHEP {\bf 1401} (2014) 088
  [arXiv:1311.0305 [hep-th]].
  \end{thebibliography}
  \end{document}